\documentstyle[twocolumn,prl,aps,epsfig]{revtex}
\tightenlines

\def\CuTi{CaCu$_{3}$Ti$_{4}$O$_{12}$ }
\def\cm-1{cm$^{-1}$}

\begin{document}
\twocolumn[
\hsize\textwidth\columnwidth\hsize\csname@twocolumnfalse\endcsname
\draft
\title
{Antiferromagnetism in \CuTi studied by magnetic Raman
spectroscopy
}

\author{
A. Koitzsch$^{1}$, G.~Blumberg$^{1,\dag}$, A. Gozar$^{1}$,  B.
Dennis$^{1}$, A.P. Ramirez$^{1,2}$, S. Trebst$^{1}$, and
Shuichi Wakimoto$^{3,\ddag}$}

\address{
$^{1}$Bell Laboratories, Lucent Technologies, Murray Hill, NJ 07974 \\
$^{2}$Los Alamos National Laboratory, Los Alamos NM, 87545 \\
$^{3}$Brookhaven National Laboratory, Upton, NY and \\
Massachusetts Institute of Technology, Cambridge, MA 02139
}

\date{13 March 2001, 
To be published in Phys. Rev. B, 01 February 2002}
\maketitle

\begin{abstract}

For \CuTi -- an insulator that exhibits a giant dielectric
response above 100 K -- Cu$^{2+}$ antiferromagnetic spin ordering has
been investigated by magnetic Raman scattering and magnetization measurements.
Below the N\'eel temperature, T$_{N} = 25$~K, magnetic excitations
have been identified.
Above T$_{N}$ Raman spectra reveal short-range antiferromagnetic fluctuations
that increase with cooling like T$^{-1}$.
No deviations from Curie-Weiss law have been observed above T$_{N}$.
\end{abstract}

]
\narrowtext

Recently the insulating cubic compound \CuTi has attracted much 
interest because of
its giant change in the dielectric response with temperature.
This compound possesses a low-frequency dielectric constant, $\epsilon
\sim 10^{4}$, which is only weakly varying in the temperature range
100--600~K.
Below 100~K, however, there is an abrupt 100-fold reduction in the
value of $\epsilon$ but no structural phase transition which is
usually associated with a large dielectric response has been observed.
X-ray diffraction and thermodynamic data argue against an explanation in
terms of collective ordering of local dipole moments
\cite{Sub00,Ram00}.
It has been suggested that at low temperatures a redistribution of
electric charge in the unit cell occurs accompanied by strong changes in
the relaxational characteristic of dipolar fluctuations
\cite{Hom01}.
The \CuTi magnetic structure has been investigated by neutron diffraction
and susceptibility measurements \cite{Col77,Kim01}.
The system is antiferromagnetic below the N\'eel temperature $T_{N}=25$~K.

In this paper we investigate the magnetic properties of \CuTi single
crystals by Raman spectroscopy, magnetization and heat capacity
measurements.
The data reveal local antiferromagnetic fluctuations
between room temperature and T$_{N}$ 
that rapidly strengthen with cooling like T$^{-1}$.
This surprising result cannot be explained by frustration effects
since the magnetization exhibits no deviations from the Curie-Weiss law 
and the Weiss constant is very close to the T$_{N}$ value.

Figure 1 shows the crystal structure of \CuTi  \cite{Sub00}, a cubic
perovskite-like compound with space group Im3 (point group $T_{h}$)
\cite{Boc79}.
Important features of this structure are the TiO$_{6}$ octahedrons,
which are considered to be responsible for the giant electrical
polarizability. If the position of the Ti$^{4+}$ ion slightly shifts
with respect to the oxygen cage \cite{Sub00,Ram00} local dipole
moments arise.
Because of the tilted arrangement of the octahedrons the ordering of
the dipole moments is frustrated and no ferroelectric phase
transition has been observed \cite{Sub00,Ram00}.
However, the Ti-O bonds are set under tension.

\begin{figure}[t]
\centerline{
\epsfig{figure=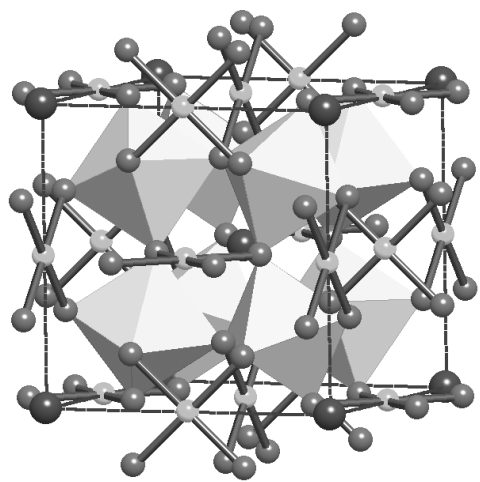,width=70mm,clip=}
}
\caption{
Crystal structure of \CuTi.
The unit cell is built up by the atoms inside the dotted lines.
It contains eight distorted TiO$_{6}$ octahedrons, copper atoms
bonded to four oxygen atoms, and large Ca atoms at the edges and in the 
middle without drawn bonds.
}
\label{Fig.1}
\end{figure}
Figure 2 shows the heat capacity and magnetization measurements that
indicate
an antiferromagnetic phase transition at 25~K with a Weiss-constant
of $\theta_{W} = -30$~K.
1/$\chi$ is linear with temperature showing Curie-Weiss
behavior almost to $T_{N}$.
The Cu$^{2+}$ ions carry a hole with spin S=1/2 in the 3d-shell,
giving rise to magnetic moments which order antiferromagnetically due 
to the superexchange interaction.
According to the Goodenough rules \cite{Goo63} it has been suggested 
\cite{Lac80} that the superexchange between the copper spins is
promoted \emph{via} the titanium ions rather than \emph{via} the usual
oxygen path.
However, the interactions are expected to be weak.

The arrangement of the Cu$^{2+}$ spins and thus the magnetic structure
is shown in Fig. 3.
We consider the effective exchange integrals $J_{1}$, $J_{2}$ and $J_{3}$
between nearest, next nearest and third nearest neighbors.
The spins are ordered antiferromagnetically along the (111) direction
with a ferromagnetic alignment in the [111] planes \cite{Kim01}.

\begin{figure}[t]
\centerline{
\epsfig{figure=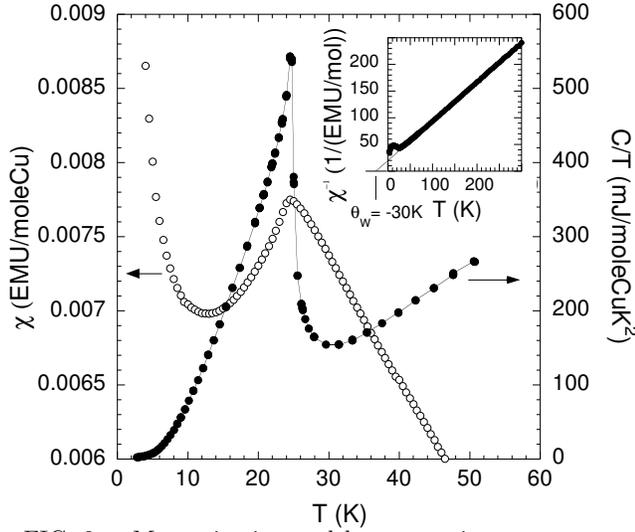,width=85mm}
}
\caption{
Magnetization and heat capacity measurements of \CuTi.
Shown are $\chi$ and C/T versus temperature.
The antiferromagnetic phase transition occurs at T$_{N} = 25$~K.
The inset shows the inverse susceptibility 1/$\chi$ versus T with a
Weiss-constant of $\theta_{W} = -30$~K.
}
\label{Fig.2}
\end{figure}

Several excitations of the magnetic ground state by photon induced
spin exchange are possible.
Direct one-magnon excitations, which correspond to the reversal of a
single spin by spin-orbit or magnetic dipole-dipole interaction, are
weak and can be identified by splitting in a magnetic field \cite{Fle68}.  
In most cases the spectra are dominated by two-magnon
excitations that correspond
to exchanging the positions of two neighboring spins.
The underlying mechanism is the following: the incoming photon
$\omega_{i}$ creates a virtual electron-hole state consisting of an
excitation across the optical gap;
then the fermions emit two-magnons at energies $\omega_{M}$ (one by
each fermion)  with momenta {\bf q} and {\bf -q} before recombining by
emitting an outgoing photon with the energy loss
$\omega_{f} = \omega_{i} - 2\omega_{M}$.
The energy loss is the result of the perturbed spin position with respect
to the ground state \cite{ShastryChubukov}.
Possible two-magnon excitations would correspond
to the exchange of the spin in position 1 with the spins in position
$1^{\prime}$ or $2^{\prime}$.
As will be shown below these processes can be resolved spectroscopically.
Furthermore the magnetic behavior above $T_{N}$ can be investigated and
compared to the magnetization measurements.

Single crystals of \CuTi were grown by the traveling-solvent floating-zone
method using a four-ellipsoidal mirror type image furnace.
Dried starting materials of CaCO$_{3}$, CuO and TiO$_{3}$ were
mixed and baked at 800$^{\circ}$C and 850$^{\circ}$C for
24 hours each with intermediate grindings to make the single-phase powder of
CaCu$_{2}$Ti$_{3}$O$_{12}$.
The feed rod was formed by a rubber tube in a hydro-static press, and
baked in air at 900$^{\circ}$C for 12 hours.
The pelletized solvent with the composition of CaCu$_{2}$Ti$_{3}$O$_{4}$ :
TiO$_{2}$ = 80 : 20 in molar ratio was placed between the feed rod
attached to the upper shaft and a seed attached to the lower shaft 
and was melted at the focal point.
Each growth was performed for $\sim 8$ hours with the growth speed of
6~mm/hour in an oxygen atmosphere.
During the growth, the upper and lower shafts rotated at 30 rpm in
opposite directions of each other to mix the molten zone.
The typical crystal size was 5~mm in diameter and 20~mm in length.

\begin{figure}[t]
\centerline{
\epsfig{figure=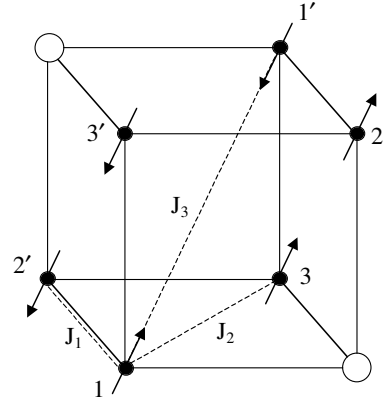,width=50mm}
}
\caption{
Spin structure of \CuTi.
The copper (black) and calcium (white) atoms of
the lower right front eighth of the unit cell (Fig. 1) are shown.
The TiO$_{6}$ octahedron is neglected for clarity.
Each Cu$^{2+}$ ion carries a spin, pointing in the (111) or
$(\bar{1}\bar{1}\bar{1})$ direction.
$J_{1}$, $J_{2}$ and $J_{3}$ are the exchange
integrals  between nearest, next nearest and third nearest neighbors.
}
\label{Fig.3}
\end{figure}

Raman measurements were performed from the [100] surface of the
crystal mounted in a continuous helium flow optical cryostat.
All spectra were taken in a backscattering geometry with excitations
from a Kr$^{+}$ laser.
The laser power was less than 10 mW and was focused to a 50 $\mu$m
diameter spot on the sample surface.
The spectra were analyzed by a custom triple grating spectrometer.
All spectra were corrected for the spectral response of the
spectrometer and detector.
The temperatures were corrected for laser heating.

In Figure 4 we present the temperature dependence of the Raman response
of  \CuTi in parallel $[z(xx)z]$ and cross $[z(xy)z]$ polarizations for
blue excitation with $\omega_{i}=2.6$~eV.
The intense peaks at high-frequency are identified as phonons.
  From the polarization dependence shown in Fig. 4 and spectra
taken for $[z(x+y,x+y)z]$ and $[z(x+y,x-y)z]$ polarizations after
sample rotation by 45$^{o}$ we conclude
that the 500 \cm-1 phonon in cross polarization has $T_{g}$ symmetry,
the phonons  at 450 and 520~\cm-1 in parallel polarization obey
$A_{g}$ symmetry and the phonon at 580 and 300~\cm-1 have $E_{g}$ symmetry.

With cooling a new feature develops at low frequencies for the
$[z(xy)z]$ polarization with a broad peak forming above T$_{N}$ and a
sharp structured band below T$_{N}$, suggesting a magnetic origin for
these excitations.
To exclude a possible one magnon origin we applied an 
8~T magnetic field. 
No shift or splitting of the peak was 
\begin{figure}[t]
\centerline{
\epsfig{figure=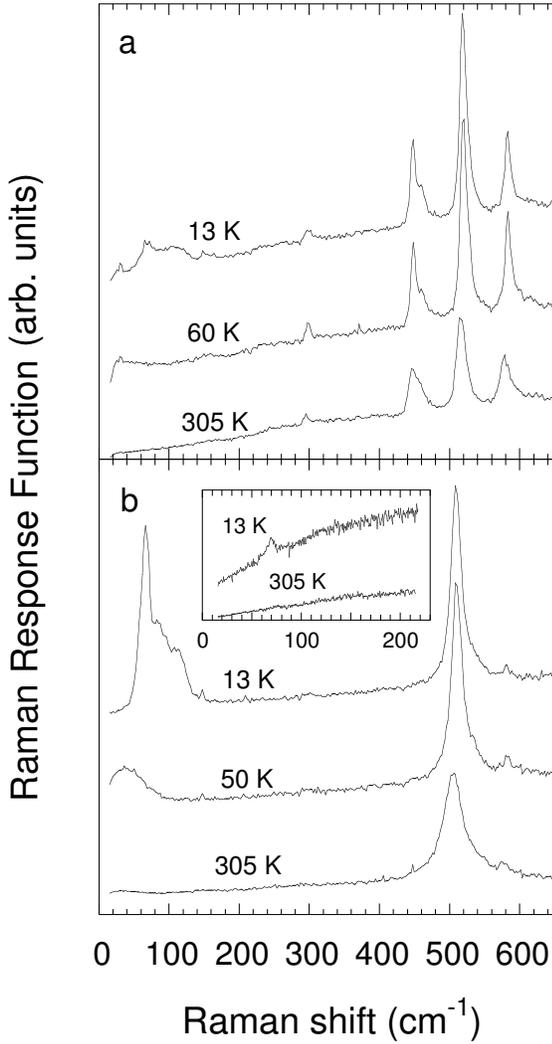,width=75mm}
}
\caption{
Raman spectra in parallel (a) and cross (b) polarization for different
temperatures using a excitation laser wavelength of $\lambda = 
482.5$~nm.
The peaks at 450, 520 and 580 \cm-1 in the parallel polarization
and 500~\cm-1 in cross polarization correspond to phonons.
The low-frequency band in cross polarization corresponds to two-magnon 
excitations.  
The inset shows spectra for excitation with $\lambda = 752$~nm  
in cross polarization. 
The magnon intensity is  suppressed indicating resonant behavior.
}
\label{Fig.4}
\end{figure}
\noindent
observed in the field 
confirming its singlet two-magnon nature.
In parallel polarization the peak is strongly suppressed.
Its polarization dependence reveals $T_{g}$ symmetry,
typical for two-magnon excitations.
The inset in Fig. 4b shows the low frequency part of the spectra
with red excitation $\omega_{i} = 1.65$~ eV for which
the two-magnon intensity is suppressed.
This demonstrates the resonance behavior of the two-magnon excitation.
The magnon intensity is expected to be sensitive to the excitation
energy because the photon-induced spin exchange process involves a
virtual electron-hole excitation to a higher energy level
\cite{ShastryChubukov}.
If the excitation energy is close to interband transitions a 
resonance enhancement occurs.

\begin{figure}[t]
\epsfig{figure=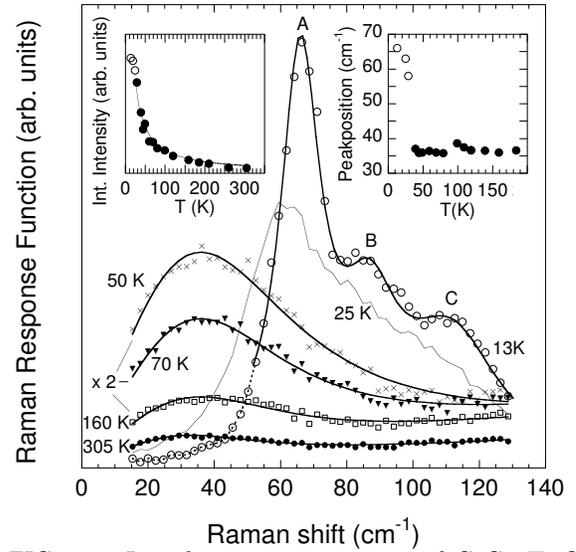,width=75mm}
\caption{
Low frequency excitations of \CuTi  in cross polarization.
Below T$_{N}$ the excitation consists of  a clear peak (A) at 66 \cm-1
and two shoulders (B and C) at  87 and 112 \cm-1.
The spectra above T$_{N}$ are fitted with a damped harmonic oscillator
function and are enlarged by a factor of two for comparison
with the magnon peak.
The 13~K spectra is fitted with three lorentzians.
The left inset shows the integrated intensity of the magnetic
excitations above
T$_{N}$ (filled circles) and below T$_{N}$ (empty circles) fitted with
$I(T) \sim T^{-1}$.
The right inset shows the peak positions of the excitations above
T$_{N}$ and peak A below T$_{N}$. }
\label{Fig.5}
\end{figure} 

Figure 5 shows the low frequency region of the Raman response.
The spectral shape depends strongly on the temperature.
Four temperature regions are distinguishable.
Above 70~K the spectra are basically flat and show weak features.
Only small deviations from the background are visible.
Below 70~K a clear peak develops around 35 \cm-1.
The peak has a rather broad shape and its intensity grows with
decreasing  temperature.
At 25~K, the value of the antiferromagnetic phase transition, the peak
shifts to higher energies and the shape becomes more structured.
Below 25~K a sharp structured two-magnon band with three components A,
B and C at 66, 87  and 112 \cm-1 is observed.
The two-magnon excitation has been fitted with 3 lorentzians between
50 and 130~\cm-1 (solid line).
The left inset shows the intensity development of the magnetic
excitation above and below T$_{N}$ with a  
linear background intensity subtracted.
The intensity increases monotonically with decreasing temperature and 
the antiferromagnetic phase transition does not immediately alter the
slope.  
The fit shows that the intensity of the fluctuations (above T$_{N}$)
is proportional to $T^{-1}$.
Note that the intensity enhancement occurs in the temperature region of
the anomalous T-dependence of the dielectric constant.
The peak positions in the right inset are constant above
the N\'eel temperature and increase suddenly around 30 K.
A slight anomaly is observed around 100 K.

We explain the three components of the two-magnon band below 25 K
with the different spin exchange processes described above.
We suggest that the two excitations at 66 and 87 \cm-1 arise from
the spin exchanges in positions 1 and $2^{\prime}$
(along $J_{1}$) and 1 and $1^{\prime}$  (along $J_{3}$).
The exchange along $J_{2}$ gives no effect.
Thus the highest excitation at 112 \cm-1 could be a four-magnon process
with instantaneous exchanges of two neighboring spin pairs.
To estimate $J_{i}$ ($i=1, 2, 3$) we
calculated the spin exchange energies
by counting the interaction energies given by
\begin{equation}
H = \sum_{<k,l>} J_{i}^{kl}({\bf S}_k \cdot {\bf S}_l)
\label{Heisenberg}
\end{equation}
where ${\bf S}_k$ and ${\bf S}_l$ are the vectors of neighboring
spins and the sum is over nearest neighbor, next nearest
neighbor and third nearest neighbor sites ($i=1, 2, 3$
correspondingly).
The energy of the two-magnon components correspond to the difference between
the total magnetic energy calculated by (1) before and after the spin
exchange.
Thus we yield a set of three linear equations to
combine with three excitation energies. The highest energy must
correspond to the assumed four-magnon process, reducing the
number of solutions to two.
One solution gives ferromagnetic exchange (and is therefore
unphysical) the second gives antiferromagnetic exchange with
$J_{1}$=23,  $J_{2}$=3 and $J_{3}$=2 \cm-1.
According to this solution we assign feature A in Fig. 5 to the 
$1 \leftrightarrow 2^{\prime}$ spin exchange,
feature B to $1\leftrightarrow 1^{\prime}$  and feature C to
$1 \leftrightarrow 2^{\prime}$ and
$2 \leftrightarrow 1^{\prime}$ (a double spin exchange).
Note that the
double spin exchange $1 \leftrightarrow 2^{\prime}$  and $2
\leftrightarrow 3^{\prime}$ yields similar results. 
The Weiss constant is approximately given by $k_{B}\theta_{W} 
=-J_{1}+2J_{2}-2J_{3}$, thus $\theta_{W}=-31$~K is in good agreement 
with the experiment.  
Our value of  $J_{1}$=23 \cm-1 is somewhat smaller than $J_{1}$=34 
\cm-1 estimated in \cite{Kim01} and we obtain $J_{1} >> J_{2}, J_{3}$ 
instead of $J_{3} >> J_{1}$, $J_{2}$ as predicted in \cite{Lac80}. 
However our minimal model has no adjustable parameters and the result is 
consistent with the magnetization measurements.

Above the N\'eel temperature no magnon excitation is expected.
We ascribe the peak around 35 \cm-1 to local antiferromagnetic spin
fluctuations, e.g. to excitations of
locally ordered spins, which we identify as overdamped magnons.
Above T$_{N}$ the long-range magnetic order is thermally destroyed.
Nevertheless, on a short-range scale spin
fluctuations persist and can be excited by light.
The fluctuations reduce the number of spins that
respond to an external field and contribute to the susceptibility.
Since the intensity of the overdamped magnons measured by Raman increases
with cooling a deviation from linearity of the
inverse susceptibility is expected.
In spite of that the susceptibility is in agreement with the
Curie-Weiss law almost down to T$_{N}$ as can be seen in Fig. 2.
Therefore we conclude that 
the fluctuations are emphasized by the resonant Raman response.

The microscopic origin of the T$^{-1}$ dependence of the intensity of the
fluctuations is not obvious.
We suggest that the proposed charge redistribution in the unit cell at
low temperatures which increases the ionicity of the bonds may enhance
the superexchange interaction and hence the intensity of the fluctuations. 
On the other hand when the temperature is increased the giant dielectric 
response is observed and the intensity of the fluctuations drops. 
Assuming that the increase in $\epsilon$ is due to a displacement of the 
Ti$^{4+}$ inside the TiO$_{6}$ octahedra one expects the fluctuations 
be disturbed if Ti$^{4+}$ is essential for the superexchange 
interaction as is proposed by \cite{Lac80}.
In this sense the increase of $\epsilon$ with 
temperature is accompanied by a decrease of the fluctuations.  

In conclusion, we have investigated the low frequency magnetic
excitations of \CuTi using Raman spectroscopy.
Below the N\'eel  temperature T$_{N} = 25$~K magnetic excitations are
identified as a three component two-magnon band.
The exchange integrals  are estimated within
the framework of the Heisenberg interaction model.
Short-range magnetic fluctuations well above T$_{N}$ were observed to
develop with cooling as T$^{-1}$. 
A possible connection between dielectric and magnetic behavior is 
discussed. 

\bigskip

We acknowledge discussions with A.M. Sengupta, S.M. Shapiro
and  M.A. Subramanian.
This work was supported in part by the Studienstiftung des Deutschen
Volkes.

\end{document}